\documentclass[times,twocolumn,final]{elsarticle}
\usepackage{medima}
\usepackage{framed,multirow}

\usepackage{amssymb}
\usepackage{latexsym}

\usepackage{url}
\usepackage{xcolor}
\usepackage{amsmath}
\usepackage{hyperref}
\usepackage{tabularx,colortbl}

\definecolor{newcolor}{rgb}{.8,.349,.1}
\newcommand{\blue}[1]{\textcolor{black}{#1}}

\journal{Medical Image Analysis}

\begin{document}

\verso{Given-name Surname \textit{et~al.}}

\begin{frontmatter}

\title{Large-Scale Multi-Center CT and MRI Segmentation of Pancreas with Deep Learning}%
%\tnotetext[tnote1]{This is an example for title footnote coding.}

\author[1]{Zheyuan Zhang}
\author[1]{Elif Keles}
\author[1]{Gorkem Durak}
\author[2]{Yavuz Taktak}
\author[1]{Onkar Susladkar}
\author[1]{Vandan Gorade}
\author[1]{Debesh Jha}
\author[2]{Asli C. Ormeci}
\author[1,2]{Alpay Medetalibeyoglu}
\author[1]{Lanhong Yao}
\author[1]{Bin Wang}
\author[1,3]{Ilkin Sevgi Isler}
\author[1]{Linkai Peng}
\author[1]{Hongyi Pan}
\author[1]{Camila Lopes Vendrami}
\author[1]{Amir Bourhani}
\author[1]{Yury Velichko}
\author[4]{Boqing Gong}
\author[5]{Concetto Spampinato}
\author[6]{Ayis Pyrros}
\author[7]{Pallavi Tiwari}
\author[8,9]{Derk C F Klatte}
\author[8,9]{Megan Engels}
\author[8,9]{Sanne Hoogenboom}
\author[9]{Candice W. Bolan}
\author[10]{Emil Agarunov}
\author[11]{Nassier Harfouch}
\author[11]{Chenchan Huang}
\author[12]{Marco J Bruno}
\author[13]{Ivo Schoots}
\author[14]{Rajesh N Keswani}
\author[1]{Frank H Miller}
\author[10]{Tamas Gonda}
\author[15]{Cemal Yazici}
\author[16]{Temel Tirkes}
\author[17]{Baris Turkbey}
\author[18]{Michael B Wallace}
\author[1]{Ulas Bagci\corref{cor1}}
\cortext[cor1]{Corresponding author}
\ead{ulasbagci@gmail.com}

\address[1]{Machine \& Hybrid Intelligence Lab, Department of Radiology, Northwestern University, Chicago, USA;}
\address[2]{Department of Internal Medicine, Istanbul University Faculty of Medicine, Istanbul, Turkey;}
\address[3]{Department of Computer Science, University of Central Florida, Florida, FL, USA;}
\address[4]{Google Research, Seattle, WA, USA;}
\address[5]{University of Catania, Catania, Italy;}
\address[6]{Department of Radiology, Duly Health and Care and Department of Biomedical and Health Information Sciences, University of Illinois Chicago, Chicago, IL, USA;}
\address[7]{Dept of Biomedical Engineering, University of Wisconsin-Madison, WI, USA;}
\address[8]{Department of Gastroenterology and Hepatology, Amsterdam Gastroenterology and Metabolism, Amsterdam UMC, University of Amsterdam, Netherlands;}
\address[9]{Department of Radiology, Mayo Clinic, Jacksonville, FL, USA; }
\address[10]{Division of Gastroenterology and Hepatology, New York University, NY, USA; }
\address[11]{Department of Radiology, NYU Grossman School of Medicine, New York, NY, USA;}
\address[12]{Departments of Gastroenterology and Hepatology, Erasmus Medical Center, Rotterdam, Netherlands;}
\address[13]{Department of Radiology and Nuclear Medicine, Erasmus University Medical Center, Rotterdam, Netherlands;}
\address[14]{Departments of Gastroenterology and Hepatology, Northwestern University, IL, USA;}
\address[15]{Division of Gastroenterology and Hepatology, University of Illinois at Chicago, Chicago, IL, USA;}
\address[16]{Department of Radiology and Imaging Sciences, Indiana University School of Medicine, Indianapolis, IN, USA;}
\address[17]{Molecular Imaging Branch, National Cancer Institute, National Institutes of Health, Bethesda, MD, USA;}
\address[18]{Division of Gastroenterology and Hepatology, Mayo Clinic in Florida, Jacksonville, USA;}

\received{xx}
\finalform{xx}
\accepted{xx}
\availableonline{xx}
\communicated{xx}

\begin{abstract}
Automated volumetric segmentation of the pancreas on cross-sectional imaging is needed for diagnosis and follow-up of pancreatic diseases. While CT-based pancreatic segmentation is more established, MRI-based segmentation methods are understudied, largely due to a lack of publicly available datasets, benchmarking research efforts, and domain-specific deep learning methods. In this retrospective study, we collected a large dataset (767 scans from 499 participants) of T1-weighted (T1W) and T2-weighted (T2W) abdominal MRI series from five centers between March 2004 and November 2022. We also collected CT scans of 1,350 patients from publicly available sources for benchmarking purposes. We introduced a new pancreas segmentation method, called \textit{PanSegNet}, combining the strengths of \textit{nnUNet} and a \textit{Transformer} network with a new linear attention module enabling volumetric computation. We tested \textit{PanSegNet}’s accuracy in cross-modality (a total of 2,117 scans) and cross-center settings with Dice and Hausdorff distance (HD95) evaluation metrics. We used Cohen’s kappa statistics for intra and inter-rater agreement evaluation and paired t-tests for volume and Dice comparisons, respectively. For segmentation accuracy, we achieved Dice coefficients of 88.3\% (± 7.2\%, at case level) with CT, 85.0\% (± 7.9\%) with T1W MRI, and 86.3\% (± 6.4\%) with T2W MRI. There was a high correlation for pancreas volume prediction with $R^2$ of 0.91, 0.84, and 0.85 for CT, T1W, and T2W, respectively. We found moderate inter-observer (0.624 and 0.638 for T1W and T2W MRI, respectively) and high intra-observer agreement scores. All MRI data is made available at \url{https://osf.io/kysnj/}. Our source code is available at \url{https://github.com/NUBagciLab/PaNSegNet}.
\end{abstract}


\end{frontmatter}

%\linenumbers

\section{Introduction}
\label{sec:intro}
Computed tomography (CT) imaging plays a pivotal role in the diagnosis and management of pancreatic diseases. CT is widely available, can be rapidly performed, and provides detailed cross-sectional imaging of the pancreas and surrounding structures. These benefits make CT an invaluable tool for detecting a wide range of pancreatic diseases, including acute and chronic pancreatitis, as well as cystic or solid neoplastic lesions~\citep{busireddy2014pancreatitis}. Recent breakthroughs in deep learning algorithms using CT scans have demonstrated improved accuracy in detecting pancreatic cancer, underscoring the diagnostic utility of this modality and the potential of artificial intelligence \citep{cao2023large}. 

Despite these advancements, Magnetic Resonance Imaging (MRI) offers two invaluable advantages over CT: lack of exposure to ionizing radiation and superior soft tissue contrast resolution. These benefits translate to greater diagnostic accuracy, particularly for pancreatic cystic lesions~\citep{pamuklar2005mr}. Volumetric segmentation of the pancreas on both CT and MRI plays a vital role in diagnosis and monitoring various pancreatic diseases, such as chronic pancreatitis (CP) and diabetes mellitus (DM)~\citep{scialpi2016pancreatic}. 

For example, a decrease in the size and volume of the pancreas can indicate the presence of these conditions, as these changes correlate with altered endocrine and exocrine functions. Pancreatic segmentation is also strategically critical in preoperative planning as it provides information that may improve the success rate and safety of surgical and interventional procedures. For example, in pancreatectomy (partial or total removal of the pancreas), accurate pancreas volumetry (gleaned from segmentation) can assist surgeons in understanding the complex anatomy, pancreas volume, and surrounding structures. This information is crucial for predicting surgical outcomes and minimizing potential complications. Similarly, in pancreas or islet cell transplantation, segmentation-based volumetry helps assess the size and suitability of the donor pancreas for transplantation. It can also help monitor post-transplant recovery and the integration of the transplanted tissue. 

Furthermore, segmentation is equally important for computational methods in medicine: any computer vision and artificial intelligence (AI) algorithm applied to the pancreas for detecting and analyzing abnormalities relies heavily on accurate pancreas segmentation as the initial step~\citep{cao2023large}. While segmentation is a critical task, developing automatic algorithms for this purpose has been challenging due to the complex and variable shape of the pancreas and the occasional loss of separation from surrounding structures. Achieving the required pixel-level precision for boundary determination demands significant time, labor, and expertise~\citep{cao2023large}. 

There is a growing demand for deep learning-based automated (or semi-automated) pancreatic segmentation tools. CT has been the primary modality among deep learning-based segmentation development efforts due to its availability, faster acquisition times, less motion artifacts, and fewer sequences than MRI. Progress on MRI-based segmentation has been lagging behind CT~\citep{zhang2023deep} due to a lack of benchmarks, effective deep learning-based segmentation methods, and large, high-quality publicly available MRI datasets. This knowledge gap presents an opportunity for innovation and research, driven by the unique challenges of MRI data, including artifacts, motion, and inhomogeneities. 

%A key issue in these hurdles lies in the discrepancy  between recent benchmarks in CT data, lack of benchmarks in MRI data, and unavailable algorithms to be tested in real-world clinical practice, specifically for MRI. Here, we summarize the two main challenges:

%\textit{(a) Limited Data Size:} Acquiring and annotating 3D medical data is expensive, leading to existing benchmarks containing a restricted number of data samples and organ annotations. This is a more challenging issue especially for MRI.

%\textit{(b) Lack of Diversity:} The high cost often confines data collection to a single center, scanner type, imaging phase, and disease state. Models trained on such homogeneous and limited datasets suffer from catastrophic performance drops when encountering test data distributions that differ from the training data. To bridge this gap, developing a more realistic and robust solution is crucial.

\textbf{Our contributions:} This study aims to bridge the following gaps:
\begin{itemize}
\item We develop an accurate and first-ever cross-platform compatible (T1W, T2W, and CT) pancreatic segmentation tool, named \textit{PanSegNet}.
\item \textit{PanSegNet}  is a new network incorporating innovative "linear self-attention" blocks specifically designed for pancreas segmentation. This approach addresses the unique challenges of the pancreas's complex anatomy.  
\item We created and shared a large-scale, publicly available, multi-center MRI dataset with detailed annotations for both T1W and T2W scans. This valuable resource fills a crucial gap in the field and unlocks further research opportunities.
\item \textit{PanSegNet} underwent rigorous validation, including cross-domain comparisons between CT and MRI scans. This comprehensive evaluation ensures its robust performance and generalizability via multi-center evaluation.
\item Our dataset can also be used for domain generalization benchmarking and experiments for abdominal MRI scans, as the medical imaging field does not have many publicly available datasets.
\end{itemize}

\section{Related Works}
Methods for segmentation of pancreas in the literature can be divided into two folds: the methods that are specifically developed for single-organ (pancreas) segmentation, and the methods that are developed for multi-organ segmentation and the pancreas is segmented as a part of multi-organ settings. Notably, the methods developed for multi-organ segmentation may have inferior results for pancreas segmentation than the methods developed for single-organ-based methods. Understandably, training a model focused solely on the pancreas allows for a more specialized feature extraction process, potentially leading to higher segmentation accuracy compared to a multi-organ model that needs to handle multiple organs with diverse characteristics. Further, single-organ segmentation models can be simpler in design compared to multi-organ models, potentially requiring fewer training resources and reducing computational complexity. Last but not least, training a model for a single organ might require less labeled data compared to a multi-organ model, which can be beneficial when dealing with limited datasets for the pancreas.

\textit{Single organ segmentation:} For many years, single organ segmentation has been the cornerstone of medical image analysis, with a plethora of successful solutions emerging~\citep{nnunet,hdenseunet,zhu20183d,roth2018application}. Take \textit{H-DenseUNet}~\citep{hdenseunet} for instance, which tackles precise liver and tumor segmentation by leveraging a hybrid 2D/3D network architecture for enhanced feature extraction, and is later used for pancreas segmentation too. The \textit{nnUNet}~\citep{nnunet} proposes a self-configurable framework based on the foundational \textit{UNet}~\citep{unet} and achieves superior performance in segmenting not only the liver, but also the spleen, kidney, and pancreas. This versatile approach readily adapts to multi-organ segmentation tasks too. Addressing the challenge of small and elongated target organs such as the pancreas, a series of works have adopted cascaded network structures~\citep{roth2018application,hussain2021vest,yan2021multi}. These networks progressively refine segmentation results in a coarse-to-fine manner. 

In our earlier work for pancreas segmentation from CT scans~\citep{zhang2022dynamic}, we have presented a dynamic linearization of the Transformer self-attention integrated with U-Net architecture, where our dynamic linearization layers were integrated along the skip connections between encoder and decoder layers to enhance segmentation performance. We have shown state-of-the-art results in CT pancreas segmentation. In this work, we go beyond this strategy by (i) simplifying the linearization procedure via dropping the dynamic content, (ii) replacing the \textit{U-Net} architecture with self-configurable \textit{nnUNet}, (iii) including large number of CT scans for rigorous comparisons, and (iv) developing a unique data sets of MRI (both T1W and T2W) and obtaining the state of the art results on these datasets, (v) assessing the generalization of the proposed method in multi-center settings. 

\textit{Multi-organ segmentation:} The task of simultaneously segmenting multiple organs within an image, necessitates networks with a heightened ability to differentiate pixel-wise features. \textit{OAN}~\citep{wang2019abdominal} tackles this challenge by designing a fusion network that leverages 2D multi-view images as input and reconstructs the final segmentation result in 3D. \textit{DenseVNet}~\citep{gibson2018automatic} proposes a dense 3D network architecture to achieve improved performance. Multi-organ segmentation has several advantages over single-organ segmentation. For instance, by incorporating information from surrounding organs, multi-organ segmentation models can potentially achieve more accurate pancreas segmentation, especially in cases of ambiguous boundaries or close proximity to other organs. However, this requires a lot more labeling and data, which is practically difficult. Multi-organ segmentation allows for simultaneous segmentation of multiple organs in a single scan, potentially streamlining clinical workflow and reducing analysis time. 

While the motivation and clinical significance are high compared to single-organ segmentation, data scarcity remains a hurdle in multi-organ segmentation. Several works~\citep{partial,prior} address this by proposing novel training paradigms that enable segmentation using partially labeled annotations from single-organ datasets. %Another line of research [28, 6, 7, 24] treats feature patches or image patches as tokens, facilitating efficient non-local context modeling across arbitrary positions. This approach has led to state-of-the-art performance on popular benchmarks [12]. 

In summary, the choice between single-organ and multi-organ segmentation depends on the specific application and available resources. For situations where the highest possible accuracy is paramount and sufficient labeled data is available, single-organ segmentation might be preferred as this is the case in our current study and we are aiming to focus on pancreas-specific anatomy delineation for clinical purposes. %However, for clinical settings where efficiency, workflow, and generalizability are essential, multi-organ segmentation can be a powerful tool for accurate pancreas segmentation.

%All these methods are presented along with certain datasets available in the public, or in-house generated specific datasets. It is worth to note that CT-based pancreas studies are numerous, and their numbers are being increased while MRI-based pancreas segmentation studies are highly limited. 

Table~\ref{comparison} enlists the prior studies on CT and MRI-based pancreas segmentation methods, the datasets they have used, the segmentation accuracy reported in those datasets, and their main architecture choices. Most of these CT-based segmentation methods utilized UNet style networks with some modifications (such as attention or recurrent connections)~\citep{nnunet}. They used a publicly available NIH dataset~\citep{clark2013cancer} and their maximum reported dice scores were saturated around ~86\%, with one exception reaching 88\%. However, none of these methods were verified with external validation, and therefore, their performance in real-world settings is unknown.

MRI-based segmentation literature is even more limited since available methods used in-house gathered data at a relatively small scale such that the maximum reported number of MRI scans was 79. Also, these methods used either T1W or T2W images and explored hierarchical features to enhance boundary detection of the pancreas or recurrent and contextual features for improved learning of pancreas location. Average dice scores for existing studies are around 70\%, and similar to CT-based studies, their success in external validation is unknown. 

In this study, we address the existing challenges in the pancreas segmentation literature by developing and comprehensively validating a novel deep learning-based precise pancreas segmentation method on a large-scale public CT dataset and introducing the first-ever multi-center, large-scale MRI pancreas dataset.

\begin{table*}
\centering
\caption{Existing CT (light gray) and MRI-based (gray) pancreas segmentation methods are enlisted. The table shows methods in the first column, and its basic strategy/approach in the second column. The datasets used in the approach are indicated in the third column. The last column shows the best dice score obtained by the method.}
\begin{tabularx}{\textwidth}{|X|X|X|c|} 
\hline
\textbf{Methods} & \textbf{Approach} & \textbf{Dataset} & \textbf{Performance / Dice} \\ \hline
\rowcolor[rgb]{0.95, 0.95, 0.95} Attention U-Net: Learning Where to Look for the Pancreas ~\cite{oktay2018attentionunet} & Attention mechanisms (layers) are integrated within the U-Net to focus on the pancreas region to avoid false positives. & NIH (8) & 83.1 ± 3.8 \\
\hline
\rowcolor[rgb]{0.95, 0.95, 0.95}  Fully automated pancreas segmentation with two-stage 3D convolutional neural networks ~\cite{zhao2019fully} & A two-stage 3D model is designed with the first stage for coarse pancreas segmentation and the second stage for refined segmentation. & NIH (8) & 86.0 ± 4.5 \\ \hline
\rowcolor[rgb]{0.95, 0.95, 0.95} Automated pancreas segmentation and volumetry using deep neural network on computed tomography ~\cite{lim2022automated} & This paper performs four individual three-dimensional pancreas segmentation networks on 1006 participants. & 1006 in-house CT scans & 84.2 \\ \hline
\rowcolor[rgb]{0.95, 0.95, 0.95} Automated pancreas segmentation using recurrent adversarial learning ~\cite{ning2018automated} & A recurrent adversarial learning framework is developed to enhance the pancreas segmentation robustness. & NIH (8) & 88.72 ± 3.23 \\ \hline
\rowcolor[rgb]{0.95, 0.95, 0.95} Deep Q-learning-driven CT pancreas segmentation with geometry-aware U-Net ~\cite{man2019deep} & A combination of deep Q-network and geometry-aware U-Net introduce reinforcement learning to improve the pancreas segmentation performance further. & NIH (8) & 86.9 ± 4.9\\ \hline
\rowcolor[rgb]{0.9, 0.9, 0.9}  Pancreas segmentation in MRI using graph-based decision fusion on convolutional neural networks ~\cite{cai2016pancreas} & The paper conducts pancreatic detection with spatial intensity context and pancreas segmentation by graph-based decision fusion. & 78 in-house T1 MRI scans & 76.1 ± 8.7\\ \hline
\rowcolor[rgb]{0.9, 0.9, 0.9}  Hierarchical 3D Feature Learning for Pancreas Segmentation ~\cite{proietto2021hierarchical} & A multiheaded decoder structure is designed to predict intermediate segmentation maps, and the final segmentation result comes from the aggregation of each level prediction. & 40 In-house T2 scans & 77.5 ± 8.6\\ \hline
\rowcolor[rgb]{0.9, 0.9, 0.9}  Improving deep pancreas segmentation in CT and MRI images via recurrent neural contextual learning and direct loss function ~\cite{cai2017improving} & The paper proposes recurrent neural contextual learning and a direct loss function and involves training the network to learn contextual information from neighboring pixels in the image. & 79 in-house T1 MRI scans & 80.5 ± 6.7\\ \hline
\end{tabularx}
\label{comparison}
\end{table*}

\section{Materials and Methods}
\label{sec:methods}
\begin{figure*}
\centering
\includegraphics[width=1\textwidth]{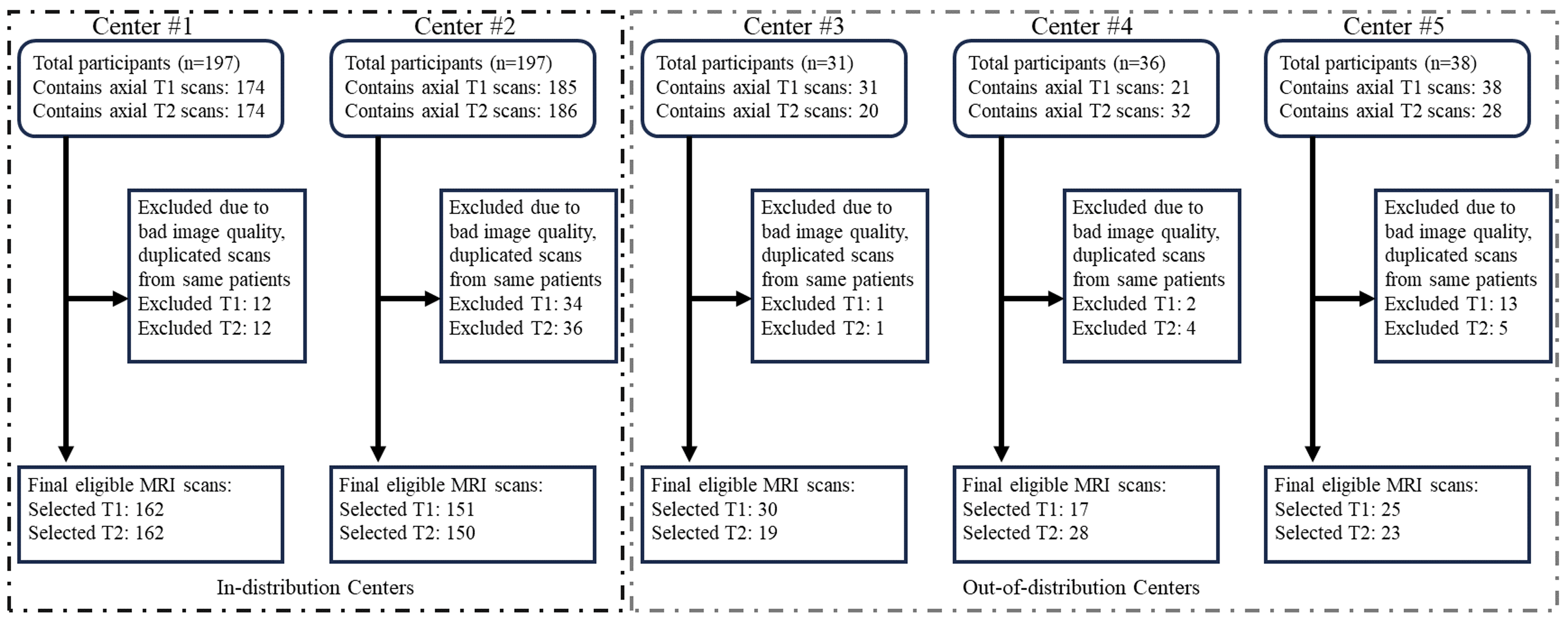}
\caption{Flowchart showing the determination of the final study population. We select Center \#1 and Center \#2 data centers as in-distribution centers (internal validation) for five cross-fold training and Center \#3, Center \#4, and Center \#5 as out-of-distribution centers (external validation). Center\#1:, Center\#2:, Center\#3:, Center\#4:, Center\#5:}
\label{datacollection}
\end{figure*}

In this IRB-approved retrospective multi-center study, 767 MRI scans from 499 adult participants from five centers were obtained using T1W and T2W imaging between March 2004 and November 2022. Scans from participants referred to MRI for pancreatic cystic lesions and suspected pancreatic adenocarcinomas were included. Figure~\ref{datacollection} illustrates our data selection procedures for multimodal MRIs. We also collected CT data from publicly available sources belonging to 1,350 adult participants who underwent CT scanning for a diverse set of indications. A total of 2,117 scans were evaluated with the presented segmentation tool, \textit{PanSegNet}.

\subsection{CT Dataset}
\textbf{Data collection:} Our study effectively utilizes several diverse, publicly available datasets:
\begin{itemize}
    \item \textbf{AMOS:} 200 scans with multi-organ segmentation including pancreas~\citep{ji2022amos}.
    \item \textbf{AbdomenCT-1K:} 1,000 scans curated from 12 centers including NIH~\cite{clark2013cancer} and MSD~\citep{antonelli2022medical}, where some of the cases are from pancreas cancer participants~\citep{ma2021abdomenct}.
    \item \textbf{WORD:} 120 healthy control scans with pancreas segmentation~\citep{luo2021word}. 
    \item \textbf{BTCV:} 30 healthy control scans with pancreas segmentation~\citep{landman2015miccai}. 
\end{itemize}

We used the AbdomenCT-1K dataset for internal validation (cross-validation) and assessed generalizability on AMOS, WORD, and BTCV datasets (totaling 350 scans) compared to their existing ground truths (available with the corresponding datasets).

\subsection{MRI Dataset}
\textbf{Data collection:} After IRB approval, we collected 767 MRI scans (385 T1W and 382 T2W) from five institutions (Centers \#1-5, Figure~\ref{datacollection}). Both sequences were included to leverage their complementary strengths: T1W images are particularly effective for visualizing specific pancreatic pathologies like adenocarcinoma, appearing as hypointense lesions, while all T1W images were acquired in the venous phase. In contrast, cystic lesions appear more conspicuous on T2W images due to their high fluid content~\citep{minami1989cystic}.  We analyzed scans across all centers. Table~\ref{data} provides comprehensive information about patient demographics and MRI parameters associated with each scan. Centers are the following: Center\#1: New York University (NYU) Medical Center, Center\#2: Mayo Clinic Florida (MCF), Center\#3: Northwestern University (NU), Center\#4: Allegheny Health Network (AHN), Center\#5: Mayo Clinic Arizona (MCA).

\begin{table*}
\centering
\caption{Patient Demographics. The table provides an overview of the dataset composition, comprising a total of 385 T1W scans and 382 T2W scans obtained from diverse imaging devices, including GE and Siemens, across five distinct medical centers. The diversity in data sources is instrumental in enhancing the robustness and generalizability of our model, reflecting real-world clinical variations and scenarios. Center\#1: New York University (NYU) Medical Center, Center\#2: Mayo Clinic Florida (MCF), Center\#3: Northwestern University (NU), Center\#4: Allegheny Health Network (AHN), Center\#5: Mayo Clinic Arizona (MCA).}
\begin{tabular}{|c|c|c|c|c|c|} 
\hline
\rowcolor[rgb]{0.9, 0.9, 0.9} \textbf{Data centers}	& \textbf{Center 1}	& \textbf{Center 2} &	\textbf{Center 3} &	\textbf{Center 4}	& \textbf{Center 5} \\ \hline
\textbf{Imaging Device}	& Siemens, GE	& Siemens, GE	& Siemens&	N/A	& Siemens, GE\\ \hline
\textbf{MRI Magnet (T)}	& 1.5, 3	& 1.5, 3	& 1.5, 3& 	N/A	& 1.5, 3 \\ \hline
\rowcolor[rgb]{0.9, 0.9, 0.9} \multicolumn{6}{|c|}{\textbf{Demographics data for T1 modalities}} \\ \hline 
\textbf{Patient Count}	& 162&	151&	30	&17	&25 \\ \hline
\textbf{Female}	&96	&87	&15	&12	&14 \\ \hline
\textbf{Male}	&66	&64	&15	&5	&11 \\ \hline
\textbf{Median Age (y)}	&64.0	&66.0	&63.0	&N/A	&71.0 \\ \hline
\rowcolor[rgb]{0.9, 0.9, 0.9} \multicolumn{6}{|c|}{\textbf{Demographics data for T2 modalities}} \\ \hline 
\textbf{Patient Count}&	162&	150&	19&	28&	23 \\ \hline
\textbf{Female}&	96&	86&	11&	19&	10 \\ \hline
\textbf{Male}&	66&	64&	8&	9&	13 \\ \hline
\textbf{Median Age (y)}&	64.0&	66.0&	63.0&	N/A& 	69.0 \\ \hline
\end{tabular}
\label{data}
\end{table*}

\subsection{Ground truth annotations and inter-observer agreement}
Five radiologists (one per center) manually segmented the pancreas on axial T1W and T2W MRI scans using ITK-SNAP~\citep{yushkevich2016itk}. A senior radiologist double-checked the annotations for quality and consistency. We conducted inter-observer and intra-observer agreement analyses using both Dice and Cohen’s kappa scores with a senior radiologist and a junior one. A random selection of 50 MRI scans from all participating centers was used for inter-observer analysis. To determine intra-observer agreement, the same radiologists assessed 20 randomly selected MRI scans a second time after a wash-out period of four weeks. On average, it takes about 25 minutes to annotate the pancreas from scratch for an experienced radiologist. We had an annotation protocol among radiologists prior to annotations; everybody used the same software (ITKSnap) for annotation purposes, and there were no significant differences in timing among participating radiologists. Generally, manual segmentation of the pancreas is time-consuming due to the low contrast and complexity of the pancreas's structure.

\subsection{Deep learning-based pancreas segmentation} 
This work addresses two key limitations in pancreas segmentation: limited MRI data and deep learning network design for volumetric and precise segmentation of the pancreas. We address the first by creating a large, multi-center MRI dataset with T1W and T2W annotations. For the second, we introduce a new segmentation algorithm, called \textit{PanSegNet}, a specialized network for CT and MRI segmentation. Building upon the established \textit{nnUNet} framework~\citep{nnunet}, \textit{PanSegNet} incorporates crucial modifications to address the specific challenges of pancreatic segmentation: the elongated and relatively small size of the pancreas necessitates specialized approaches to capture its complex structure effectively. Transformers and self-attention mechanisms show potential in recognizing the intricate structure of the pancreas, yet significant computational demands and extensive data requirements hinder them~\citep{zhang2022dynamic}. Hence, we integrate Transformers with \textit{nnUNet} and introduce a novel "linear self-attention" block into this new architecture that strategically modifies the self-attention to address the high computational cost of standard transformers, making it efficient for volumetric segmentation. This modification approximates quadratic computations with linear complexity, significantly reducing the computational burden. 

\begin{figure*}
\centering
\includegraphics[width=0.7\textwidth]{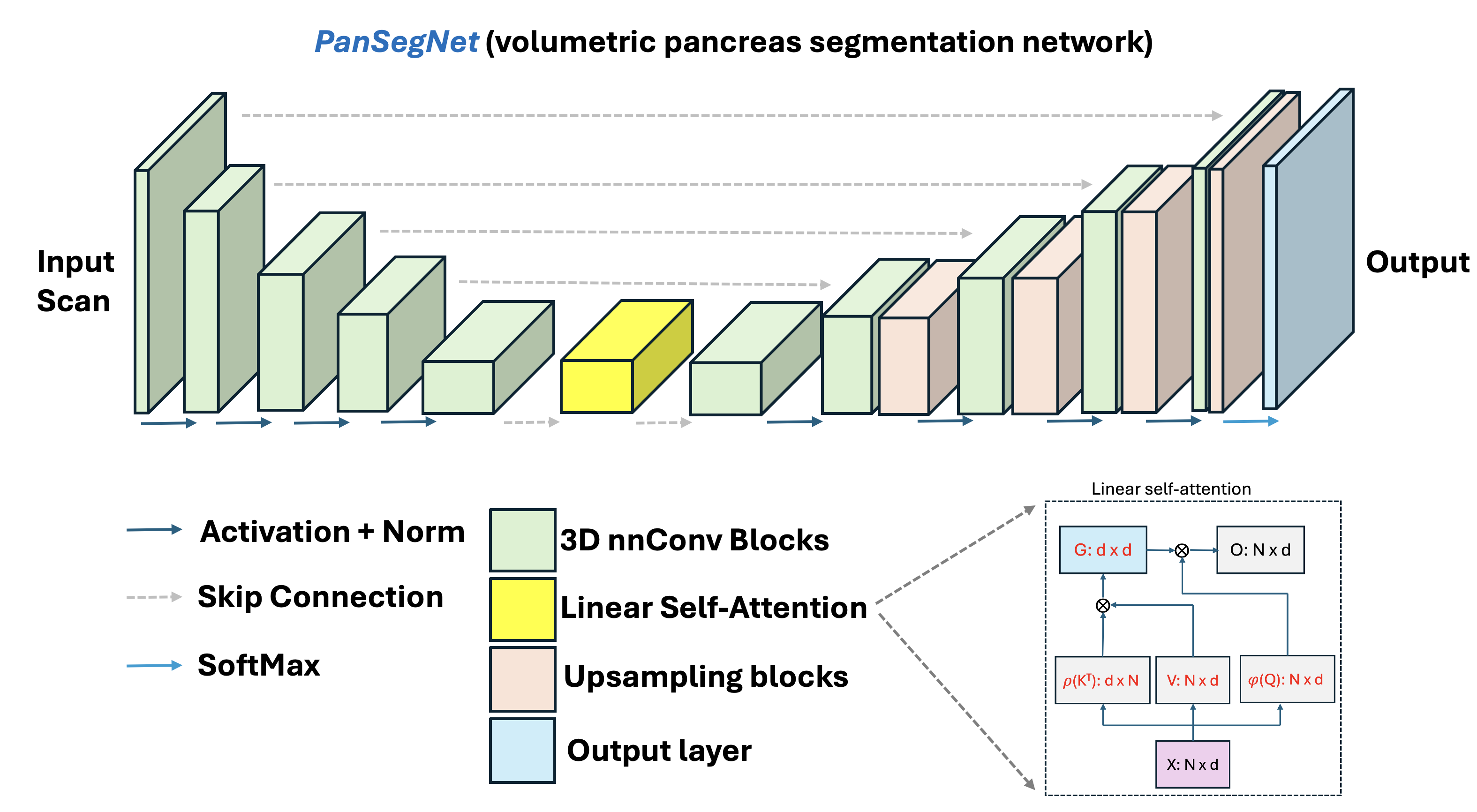}
\caption{PanSegNet is based on a combination of \textit{nnUnet} with linear self-attention mechanism. Linear self-attention is obtained by converting the self-attention mechanism with linearization operation, as described below. The architecture accepts volumetric input, therefore appreciating the full anatomy details compared to pseudo-3D approaches.}
\label{architecture}
\end{figure*}

Figure~\ref{architecture} illustrates the overall segmentation architecture of \textit{PanSegNet}, based on an encoder-decoder style segmentation where the backbone is \textit{nnUNet}, and the bottleneck is a linear self-attention layer (yellow layers) converted from traditional self-attention strategy. During the encoder process, we extract the higher-level representative features. In the decoder process, the extracted features are used to generate a segmentation mask at each hierarchy level.

\subsubsection{Linear Attention: Transformer with linear complexity}
The self-attention mechanism achieved remarkable performance in many computer vision and natural language processing tasks. However, the quadratic complexity prohibits its application in volumetric segmentation tasks. Recently some methods have been proposed to reduce the computation of self-attention to linear complexity~\citep{kitaev2020reformer,wang2020linformer,shen2021efficient}. 

Given the individual feature vectors $X \in \mathbb{R}^{n\times d}$, we have the values $V \in \mathbb{R}^{n\times d}$, the queries $Q \in \mathbb{R}^{n\times d}$, the keys $K \in \mathbb{R}^{n\times d}$ after linear projections, where $n$ represents the feature-length and $d$ represents the feature dimension. The traditional self-attention mechanism can be expressed in a more general way as follows:
\begin{equation}
    V_{i}^{'} = \sum_{j=1}^{n} sim(Q_i, K_j) V_j,
\end{equation}
where the similarity function is defined as $sim(q,k) = softmax(qk^T/\sqrt{d})$ and i, j are the index for features. In the linear Transformer paradigm, we want to replace the similarity function such that the similarity function can be divided into two separate parts using the normalized feature quantification function $sim(q, k) = \phi(q) \rho(k)^T$. This replacement could allow us to get:
\begin{equation}
    V_{i}^{'} = \sum_{j=1}^{n} (\phi(Q_i) \rho(K_j)^T) V_j= \phi(Q_i)( \sum_{j=1}^{n} \rho(K_j)^T V_j).
\end{equation}

\begin{figure}[h]
\centering
\includegraphics[width=0.5\textwidth]{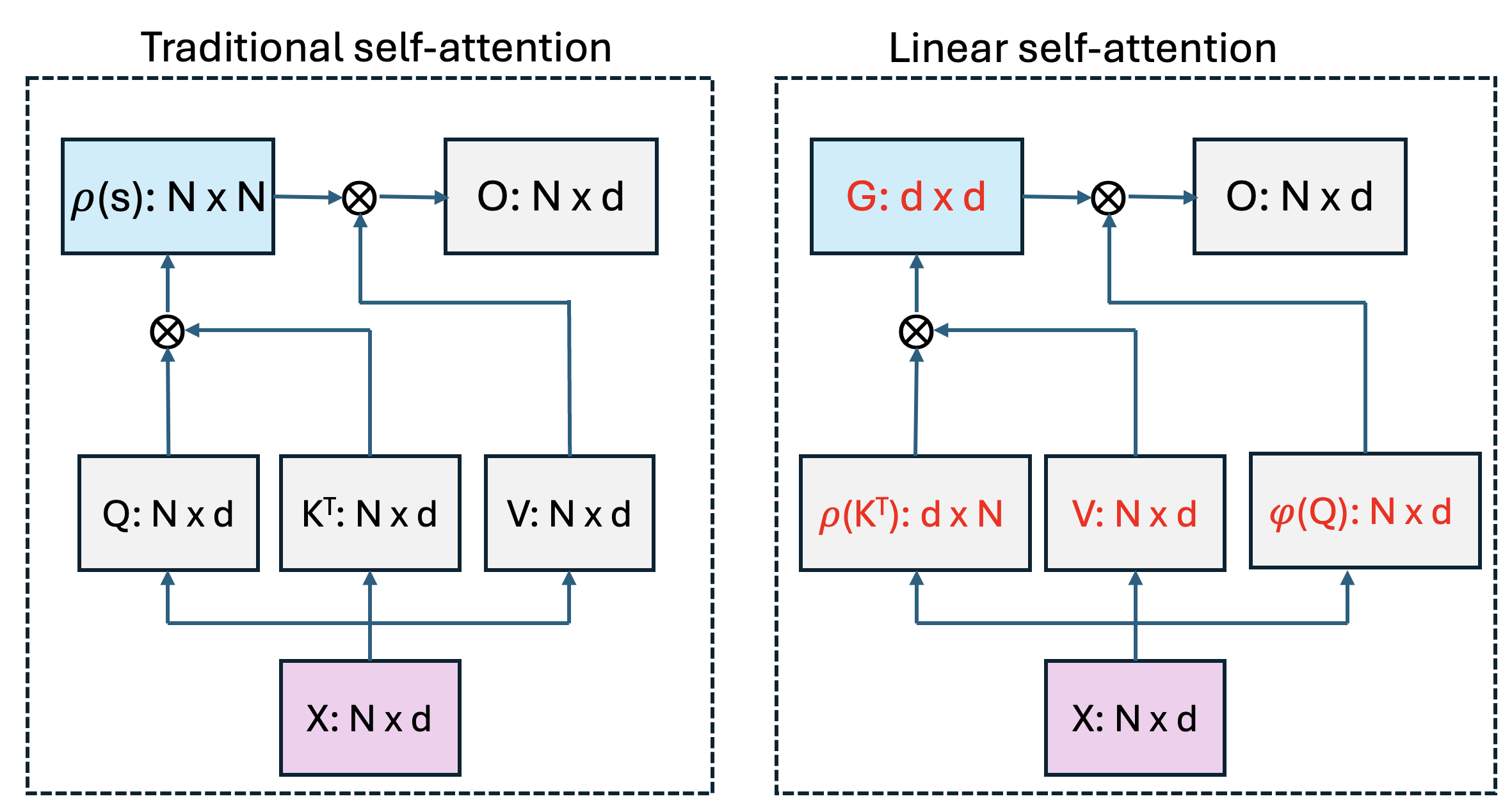}
\caption{Comparison of traditional self-attention mechanism (left) v.s. linear self-attention mechanism (right). $X$ is input, $O$ is output. Red fonts show the specific changes we apply to self-attention to linearize.}
\label{linearattention}
\end{figure}

In this paper, we employ the similarity definition of the \textit{Efficient Transformer} study~\citep{shen2021efficient}, where $\phi(Q_i), \rho(K_j)$ denote applying the \textit{SoftMax} function along each row or column of $Q, K$. This definition allows us to keep the important property of original self-attention that $\sum_{j=1}^{n} sim(Q_i, K_j) = 1$ and to reduce the complexity from quadratic to linear. Figure~\ref{linearattention} shows our proposed linear attention module (right) vs the traditional self-attention mechanism (left) of Transformers. The linear transformer paradigm seeks to reformulate the self-attention mechanism such that the computational complexity can be reduced to a linear time, $O(n)$.

We trained \textit{PanSegNet} using SGD optimization on five NVIDIA A6000 GPUs. To balance efficiency and performance, we employed a systematic hyperparameter tuning approach, iteratively adjusting the learning rate (0.01), batch size (2), and epochs (600). Dropout layers and data augmentation techniques helped prevent overfitting.

\subsubsection{Segmentation model evaluation}
We comprehensively evaluated the segmentation performance with two complementary categories of metrics: region-level metrics, including Dice score (Dice), Jaccard index (Jaccard), Precision, and Recall, and boundary-based metrics such as the 95\% Hausdorff Distance (HD95) and Average Symmetric Surface Distance (ASSD). This dual-metric paradigm thoroughly appraises the model's inherent capabilities~\citep{li2020linsem,bagci2011hierarchical}. Higher region-based metrics and lower shape-based metrics indicate superior performance.

\subsection{Multi-center setup for generalization of deep learning model}
Our comprehensive evaluation strategy includes both internal and external validation for CT and MRI modalities.

\textit{For CT Segmentation Evaluation:} We used 5-fold cross-validation on the AbdominalCT-1K dataset for internal validation and tested on AMOS, WORD, and BTCV datasets for external validation, reflecting real-world data diversity.

\textit{For MRI Segmentation Evaluation:} To ensure sufficient training data, we combined scans from Centers \#1 and \#2 for internal validation (5-fold cross-validation with T1W and T2W data). Generalizability was assessed on external data from Centers \#3, \#4, and \#5, mimicking real-world clinical variability.

\section{Results}
\subsection{Intra- and inter-observer agreement analysis}
The inter-observer agreement analysis yielded Dice scores of 80.14\% and 80.58\% for T1W and T2W MRI, respectively, highlighting the difficulty of achieving an accurate pancreas segmentation (Table~\ref{observer}). The Cohen's kappa coefficients were 0.624 and 0.638 for T1W and T2W MRI, respectively, showing moderate reliability. In contrast, the intra-observer analysis showed a high level of consistency, with Dice scores of 96\% for T1W and 93.6\% for T2W and Cohen’s kappa coefficients of 0.901 and 0.854, respectively, underscoring the stability and reliability of reference standards.

\begin{table*}
\centering
\caption{Inter-observer and intra-observer analyses show the quality and stability of pancreas segmentation annotations on the collected MRI datasets. (Note that CT data are from publicly available databases with provided annotations). }
\begin{tabular}{|c|c|c|c|c|c|c|} 
\hline
\rowcolor[rgb]{0.9, 0.9, 0.9} \multicolumn{7}{|c|}{\textbf{Multi-center T1 Pancreas Annotations}} \\ \hline
\rowcolor[rgb]{0.9, 0.9, 0.9} \textbf{Evaluations} &\textbf{Dice (\%)}&	\textbf{Jaccard (\%)}&	\textbf{Precision (\%)}&	\textbf{Recall (\%)}&	\textbf{HD 95 (mm)}&	\textbf{ASSD (mm)} \\ \hline
 \textbf{Inter-observer}&	80.14 ± 11.03&	68.34 ± 16.09&	91.15 ± 7.96&	72.54 ± 14.92&	12.23 ± 7.26&	2.17 ± 1.27 \\ \hline
\textbf{Intra-observer}&	96.00 ± 1.89&	92.38 ± 3.51&	95.85 ± 2.97&	96.24 ± 2.49&	2.84 ± 1.42&	0.38 ± 0.18 \\ \hline
\rowcolor[rgb]{0.9, 0.9, 0.9} \multicolumn{7}{|c|}{\textbf{Multi-center T2 Pancreas Annotations}} \\ \hline
\rowcolor[rgb]{0.9, 0.9, 0.9} \textbf{Evaluations} &\textbf{Dice (\%)}&	\textbf{Jaccard (\%)}&	\textbf{Precision (\%)}&	\textbf{Recall (\%)}&	\textbf{HD 95 (mm)}&	\textbf{ASSD (mm)} \\ \hline
\textbf{Inter-observer}&	80.58 ± 12.58&	69.31 ± 17.54&	97.58 ± 4.38&	70.33 ± 17.20&	12.52 ± 10.68&	2.07 ± 1.55\\ \hline
\textbf{Intra-observer}&	93.60 ± 5.30&	88.41 ± 8.95&	91.40 ± 8.44&	96.34 ± 3.75&	3.10 ± 2.68&	0.55 ± 0.53\\ \hline
\end{tabular}
\label{observer}
\end{table*}

\subsection{Segmentation performance with CT scans}
\begin{figure*}
\centering
\includegraphics[width=1\textwidth]{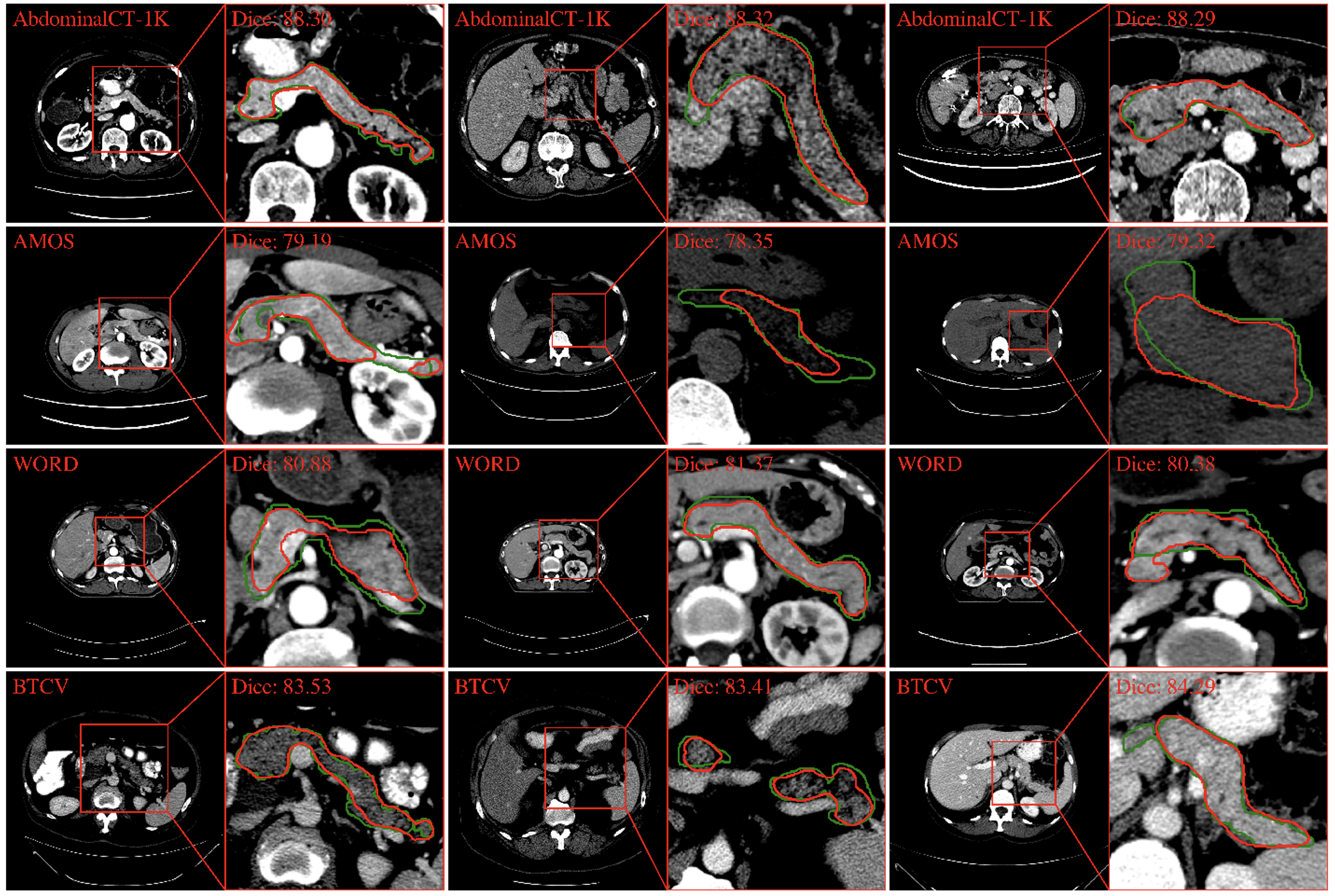}
\caption{Segmentation results for CT pancreas across multiple datasets (green indicates the predicted pancreas, and red indicates the annotations). While AbdominalCT-1K exhibits robust segmentation performance, marked by precise boundary delineation, a domain shift is observed when extending the model to the AMOS, WORD, and BTCV datasets, underscoring the significance of addressing domain shifts for clinical applications. For a fair comparison, we select the visualization samples near the median value according to the Dice coefficient distribution (note: Dice is calculated volumetrically).}
\label{ctresults}
\end{figure*}

In the AbdomenCT-1K dataset, \textit{PanSegNet} yielded an average Dice score of 88.31\% with a standard deviation of 7.24\% (median: 89.64\%) and an HD95 distance of 5.10 mm with a standard deviation of 8.43 mm (median: 3.16 mm) (Table~\ref{evaluation}). Visual examination of the segmentation results, as depicted in Figure~\ref{ctresults}, demonstrates the model's accuracy in delineating the intricate contours of the pancreas, closely aligning with the ground truth annotations. Segmentation performance dropped when our trained model was applied directly to AMOS, WORD, and BTCV external datasets. These datasets represent real-world clinical scenarios and potentially differ in distribution from the AbdomenCT-1K training data (domain shift). As expected, the model's performance drops on these datasets, indicating the limitations of directly applying models trained on specific datasets to more diverse real-world settings. Specifically, the Dice coefficients decreased to 78.79\% (-10.78\%, p-value: 2×10-32), 80.89\% (-8.40\%, 6.13×10-25), and 83.71\% (-5.21\%, p-value: 6.05×10-4), respectively. These statistically significant decreases highlight the importance of addressing domain shift challenges when deploying models in real-world clinical applications. It is worth noting that despite the domain-shift challenges, \textit{PanSegNet} obtains highly promising dice scores. 

\subsection{Segmentation performance with MRI scans}
\subsubsection{T1W MRI}
\begin{figure*}
\centering
\includegraphics[width=1\textwidth]{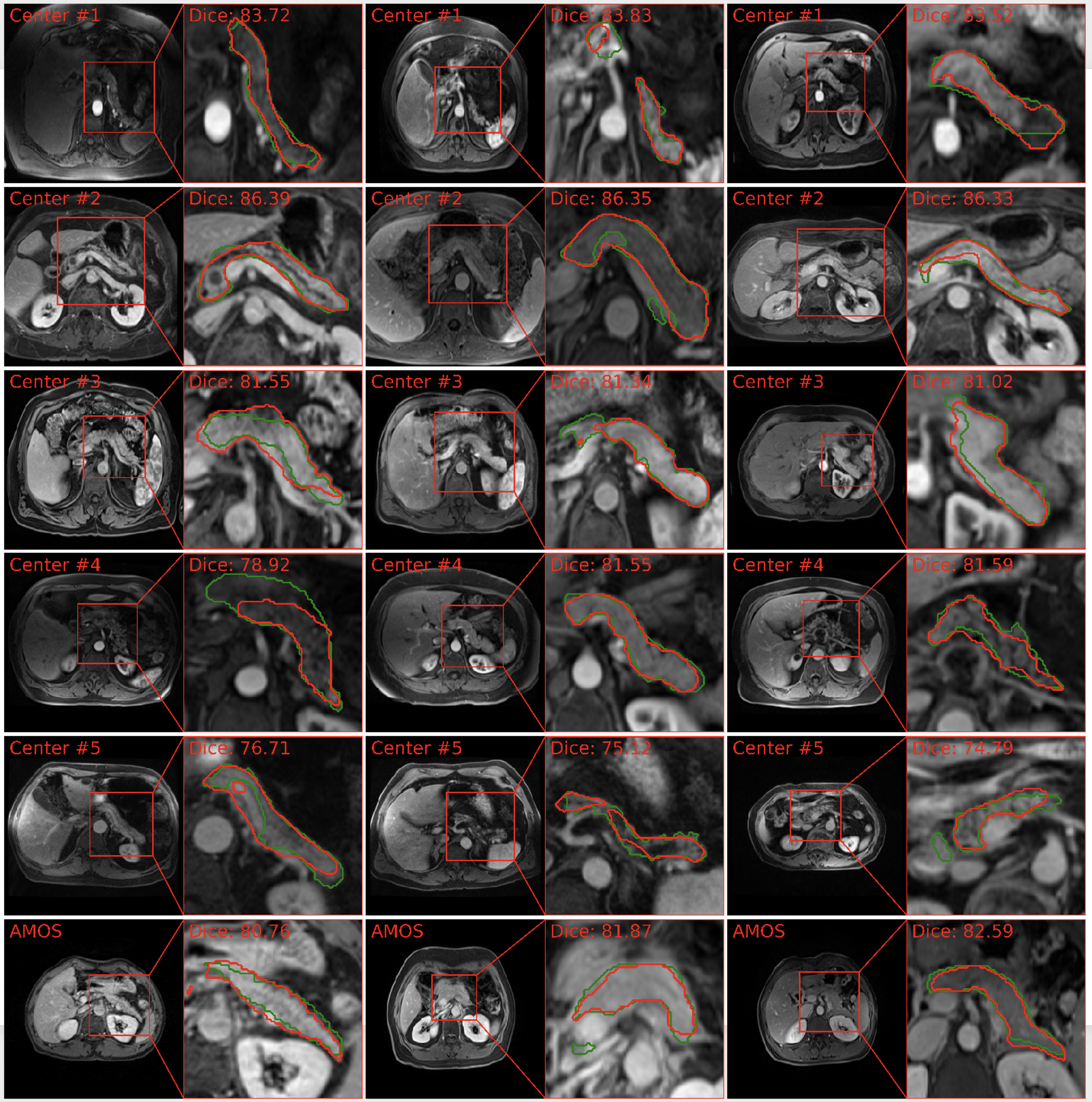}
\caption{MRI T1W pancreas segmentation visualization across various data centers. The segmentation delineations illustrate the model's capability to delineate pancreas boundaries precisely, exemplified by the accurate results. We observe domain shifts in external validation from Centers \#3, \#4, and \#5.}
\label{t1results}
\end{figure*}

For T1W MRI segmentation, our proposed \textit{PanSegNet} model exhibited noteworthy performance within the source domain, achieving average Dice scores of 85.02\% (±7.90\%, median 87.26\%) with HD95 distance of 6.37 mm (± 7.96, median 4.40 mm). Specifically, within the Center\#1 and Center\#2 datasets, our model achieved an average Dice coefficient of 86.44\% (±7.04, median: 88.89\%) and 83.70\% (±8.41\%, median: 86.01) (Table~\ref{evaluation}). Our results exhibit the model's proficiency in accurately delineating pancreatic structures within the T1W modality. This strong segmentation capability was further visually validated through our segmentation results in Figure~\ref{t1results}, which closely aligned with ground truth annotations.

\begin{table*}
\centering
\caption{Quantitative segmentation performance evaluation across various CT and MRI datasets. The row with $*$ indicates the performance in the training center using a 5-fold cross-validating (internal validation), while the rest of the rows $\dagger$ indicate the performance in the test centers (external validation). }
\begin{tabular}{|c|c|c|c|c|c|c|}
\hline
\rowcolor[rgb]{0.9, 0.9, 0.9} \multicolumn{7}{|c|}{\textbf{Multi-center CT Pancreas Segmentation}} \\ \hline
\rowcolor[rgb]{0.9, 0.9, 0.9} \textbf{Evaluations} &\textbf{Dice (\%)}&	\textbf{Jaccard (\%)}&	\textbf{Precision (\%)}&	\textbf{Recall (\%)}&	\textbf{HD 95 (mm)}&	\textbf{ASSD (mm)} \\ \hline
\textbf{AbdominalCT-1K*}&	88.31 ± 7.24&	79.71 ± 10.02&	87.77 ± 9.29&	90.08 ± 8.86&	5.10 ± 8.43&	1.17 ± 1.49\\ \hline
\textbf{AMOS$\dagger$}&	78.79 ± 18.56&	67.96 ± 19.62&	80.39 ± 17.71&	80.37 ± 19.11&	13.47 ± 22.97&	2.92 ± 5.43\\ \hline
\textbf{WORD$\dagger$}&	80.89 ± 7.48&	68.51 ± 9.60&	85.47 ± 12.46&	78.17 ± 6.77&	12.85 ± 32.46&	2.75 ± 5.90\\ \hline
\textbf{BTCV$\dagger$}&	83.71 ± 6.09&	72.43 ± 8.46&	81.84 ± 7.31&	86.17 ± 7.32&	8.29 ± 9.52&	1.59 ± 1.25\\ \hline
\rowcolor[rgb]{0.9, 0.9, 0.9} \multicolumn{7}{|c|}{\textbf{Multi-center T1 MRI Pancreas Segmentation}} \\ \hline
\rowcolor[rgb]{0.9, 0.9, 0.9} \textbf{Evaluations} &\textbf{Dice (\%)}&	\textbf{Jaccard (\%)}&	\textbf{Precision (\%)}&	\textbf{Recall (\%)}&	\textbf{HD 95 (mm)}&	\textbf{ASSD (mm)} \\ \hline
\textbf{Center \#1*}&	83.70 ± 8.41&	72.74 ± 10.76&	84.85 ± 9.10&	83.67 ± 9.71&	6.79 ± 8.95&	1.42 ± 1.51 \\ \hline
\textbf{Center \#2*}&	86.44 ± 7.04&	76.71 ± 9.76&	87.92 ± 8.04&	85.87 ± 8.51&	5.91 ± 6.71&	1.22 ± 1.19 \\ \hline
\textbf{Combined*}&	85.02 ± 7.90&	74.65 ± 10.48&	86.33 ± 8.74&	84.73 ± 9.22&	6.37 ± 7.96&	1.32 ± 1.37 \\ \hline
\textbf{Center \#3$\dagger$}&	81.55 ± 8.12&	69.49 ± 9.35&	74.44 ± 9.14&	90.55 ± 7.30&	5.64 ± 4.96&	1.80 ± 1.02 \\ \hline
\textbf{Center \#4$\dagger$}&	79.80 ± 6.15&	66.81 ± 8.17&	86.33 ± 8.78&	75.08 ± 8.13&	8.01 ± 3.53&	1.67 ± 0.82 \\ \hline
\textbf{Center \#5$\dagger$}&	76.21 ± 8.74&	62.34 ± 11.09&	73.98 ± 12.95&	80.73 ± 9.11&	14.26 ± 29.85&	2.98 ± 3.69 \\ \hline
\textbf{AMOS$\dagger$}&	81.14 ± 13.41&	69.78 ± 13.72&	82.28 ± 10.41&	80.90 ± 15.04&	12.09 ± 26.37&	2.54 ± 2.96\\ \hline
\rowcolor[rgb]{0.9, 0.9, 0.9} \multicolumn{7}{|c|}{\textbf{Multi-center T2 MRI Pancreas Segmentation}} \\ \hline
\rowcolor[rgb]{0.9, 0.9, 0.9} \textbf{Evaluations} &\textbf{Dice (\%)}&	\textbf{Jaccard (\%)}&	\textbf{Precision (\%)}&	\textbf{Recall (\%)}&	\textbf{HD 95 (mm)}&	\textbf{ASSD (mm)} \\ \hline
\textbf{Center \#1*}&	85.89 ± 5.31&	75.62 ± 7.63&	87.47 ± 6.42&	85.01 ± 7.70&	5.76 ± 4.61&	1.12 ± 0.94\\ \hline
\textbf{Center \#2*}&	86.69 ± 7.45&	77.15 ± 9.86&	89.09 ± 6.17&	85.57 ± 10.67&	5.72 ± 7.40&	1.09 ± 1.23\\ \hline
\textbf{Combined*}&	86.27 ± 6.44&	76.36 ± 8.81&	88.25 ± 6.35&	85.27 ± 9.26&	5.74 ± 6.11&	1.10 ± 1.09\\ \hline
\textbf{Center \#3$\dagger$}&	87.83 ± 4.08&	78.52 ± 6.18&	83.22 ± 6.14&	93.37 ± 3.99&	5.33 ± 4.34&	0.97 ± 0.67\\ \hline
\textbf{Center \#4$\dagger$}&	81.41 ± 10.76&	69.82 ± 12.97&	86.12 ± 13.44&	78.37 ± 10.23&	9.67 ± 12.59&	1.85 ± 3.14\\ \hline
\textbf{Center \#5$\dagger$}&	83.03 ± 3.71&	71.15 ± 5.27&	90.01 ± 4.64&	77.67 ± 7.16&	6.38 ± 3.29&	1.12 ± 0.51\\ \hline
\end{tabular}
\label{evaluation}
\end{table*}

We also observed a domain shift in segmentation performance when conducting external validation after selecting the best performance model from cross-validation using in-domain data. The Dice scores decreased to 81.55\% (-4.08\%, p-value: 1.92×10-2), 79.80\% (-6.13\%, p-value: 7.96×10-3), and 76.21\% (-10.36\%, p-value: 2.04×10-8), respectively, for data from three different centers. We also performed experiments with the publicly available AMOS dataset, as it includes 40 MRI scans for the pancreas, and the Dice score also dropped to 81.14\% (-4.56\%, p-value: 8.44×10-3). While these reductions highlight the challenges posed by domain shifts, it is worth noting that our model still maintained a strong performance even in these cross-domain scenarios.

\subsubsection{T2W MRI} 
\begin{figure*}
\centering
\includegraphics[width=1\textwidth]{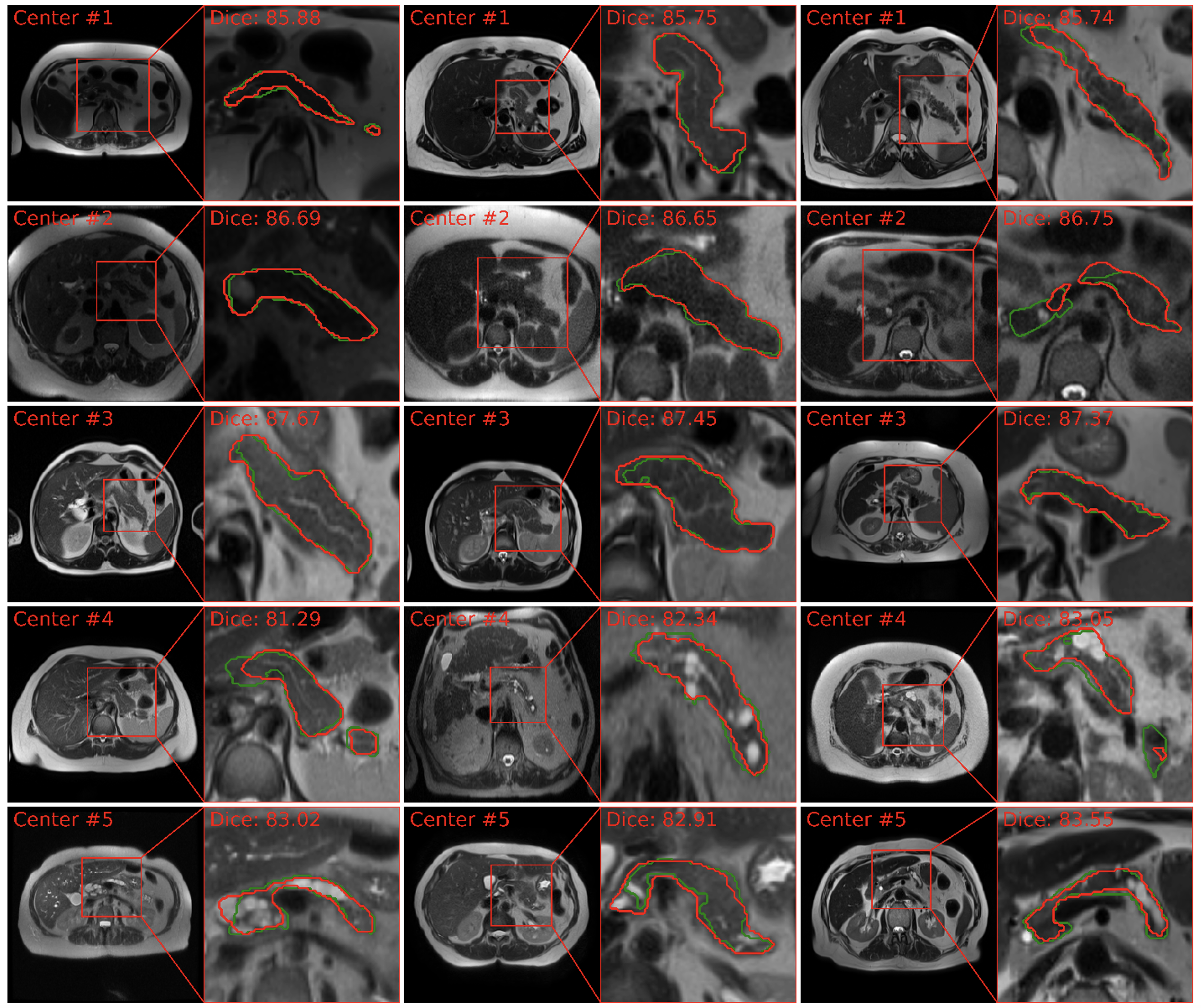}
\caption{MRI T2W pancreas segmentation visualization across various data centers. The segmentation delineations illustrate the model's capability to delineate pancreas boundaries precisely, exemplified by the accurate results. The Center \#3 T2W segmentation also exhibits relatively high results, showcasing its segmentation potential. We observe domain shifts in external validation from Centers \#3, \#4, and \#5. }
\label{t2results}
\end{figure*}

Our \textit{PanSegNet} similarly demonstrated strong segmentation capabilities on T2W sequences. Within the source domain (Center\#1 and Center\#2 datasets), the model achieved an average Dice coefficient of 86.27\% with a standard deviation of 6.44\% (median: 87.84\%). When assessing the segmentation performance in the Center\#3 dataset, we observed a considerably high Dice coefficient of 87.83\% (±1.80\%, median 89.22\%). When deploying the model in out-of-distribution settings (Center\#3 and Center\#5), the Dice coefficients were decreased to 81.41\% (-5.63\%, p-value: 4.23×10-4) and 83.03\% (-3.76\%, p-value: 1.79×10-2), respectively, implying the impact of domain shifts. Despite the shifts, the results are still at the forefront of current advancement.  Visual results are illustrated in Figure~\ref{t2results}, which are closely aligned with ground truth annotations.

\subsection{Volume and cross-modality comparisons}
\subsubsection{Pancreas volumetry:} 
\begin{figure*}
\centering
\includegraphics[width=1\textwidth]{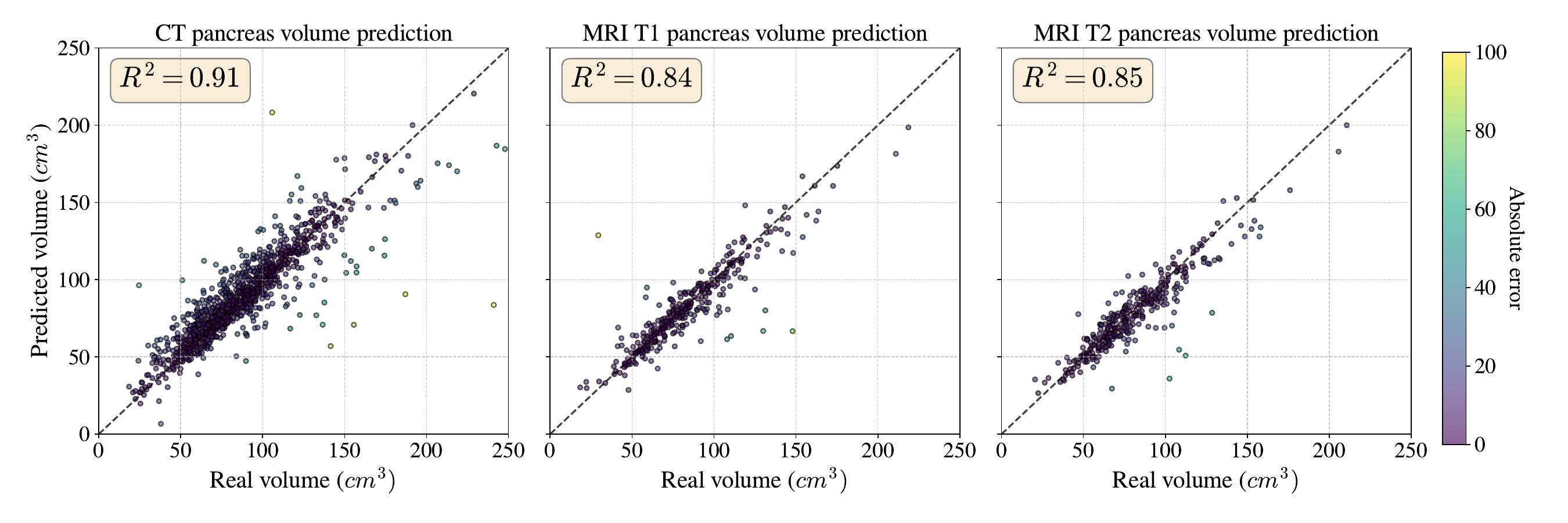}
\caption{Compelling correlation between the real volume and predicted volume for pancreas segmentation across three distinct modalities: CT, MRI T1, and MRI T2. Each subplot showcases a linear fitting line, corresponding to Pearson’s correlation $R^2$ values of 0.91, 0.84, and 0.85 for CT, MRI T1W, and MRI T2W, respectively. These high R2 values elucidate the accuracy and effectiveness of our volume prediction model, reinforcing its potential utility in clinical applications.}
\label{volumeresults}
\end{figure*}

Our volumetric comparisons (pancreas volume predicted by the \textit{PanSegNet} algorithm and ground truths) revealed significant and high correlations: each subplot in Figure~\ref{volumeresults} showcases a linear fitting line, with corresponding $R^2$ values of 0.91, 0.84, and 0.85 for CT, MRI T1W, and MRI T2W, respectively. Another direct evaluation metric for volume statistics is the absolute volume prediction error. Our methods also achieved low average absolute volume prediction error, 12.34\%, 10.49\%, and 10.05\% for CT, MRI T1, and MRI T2, respectively. These low error rates demonstrate \textit{PanSegNet}'s effectiveness in predicting pancreas volume with minimal deviation from the actual volume measurements.

\subsubsection{Visual Analysis:} Two senior radiologists independently evaluated a random selection of cases using strict same-agreement criteria. This signifies that a predicted segmentation was only considered acceptable if both radiologists concurred on its quality. Forty cases (20 T1W and 20 T2W) were selected from source Centers\#1 and \#2, and 30 cases (15 T1W and 15 T2W) were selected from target Centers \#3, \#4, and \#5. The visual acceptance rate for the source centers was 95\% (97.5\% for T1W and 92.5\% for T2W). Notably, for the external centers (Centers \#3, \#4, and \#5) whose data were not included in the training process, the model achieved a visual acceptance rate of 90\% for both T1W and T2W modalities.

\subsubsection{Cross-modality comparison:} Within the source domain, our model achieved statistically significantly higher Dice coefficients with T2W images (86.27\%) compared to T1W images (85.02\%) (p-value = 0.03). This observation was further supported by visual analysis, which revealed a generally higher acceptance rate for T2W segmentation.

\subsubsection{Generalization Results Across Centers:} 
\begin{figure*}
\centering
\includegraphics[width=1\textwidth]{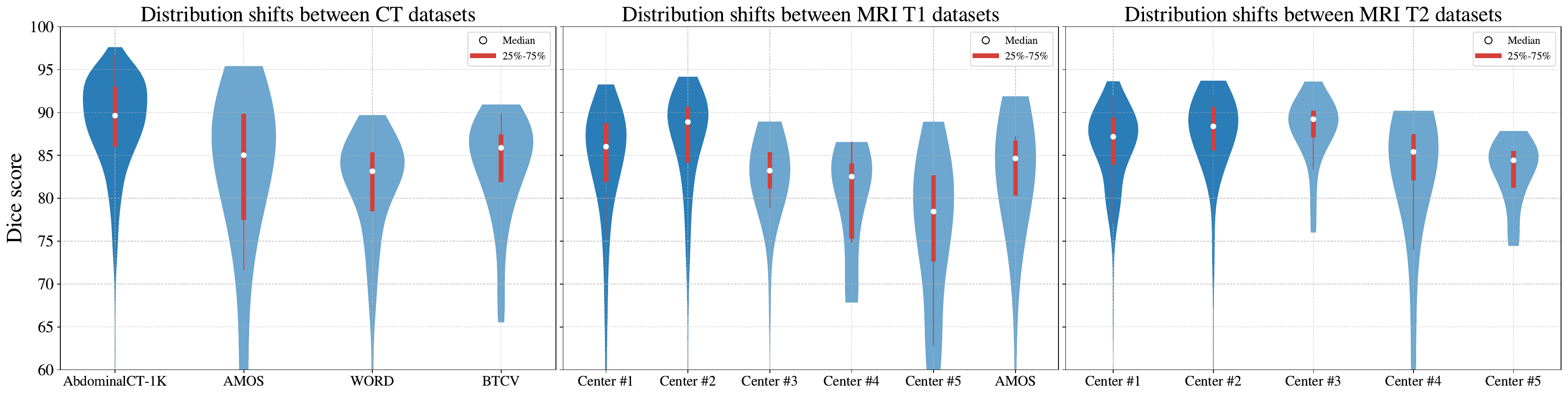}
\caption{The shifts in Dice coefficients observed across three modalities: CT, T1W, and T2W MRI scans, stemming from the influence of domain shifts. As we move from the source domain (dark blue) to other datasets (light blue), we observe variations in segmentation performance evidenced by the changing Dice coefficients.}
\label{domainshift}
\end{figure*}

As illustrated in Figure~\ref{domainshift}, a performance decrease is observed across all modalities (CT, MRI T1W, and MRI T2W) when evaluating external centers compared to source centers. This is likely attributable to domain shift, where data distribution in the external centers differs from the source centers used for training. This trend is further corroborated by the radiologists' acceptance rates, which are consistently higher for the source centers than the external centers. In our case, the domain shift wasn't severe, meaning the model could still perform its segmentation task reasonably well.

\subsection{Comparison and Benchmarking}
%\textbf{CT-based results:} 
%Here, we compare our CT segmentation results with those of other representative published work, and the results are shown in the following table.

%\begin{table*}
%\centering
%\caption{Quantitative segmentation performance evaluation with other Existing CT pancreas segmentation methods.}
%\begin{tabular}{|c|c|c|c|c|c|c|}
%\hline
%\rowcolor[rgb]{0.9, 0.9, 0.9} \textbf{Methods} & Dataset& \textbf{Dice (\%)}&	\textbf{HD 95 (mm)}&	\textbf{ASSD (mm)} \\ \hline
%Attention U-Net  & NIH & 82.1 ± 5.7 & - & 2.333 ± 0.856 \\ \hline
%29 & 1006 inhouse CT & 84.2 & - & - \\ \hline
%28 & NIH & 86.0 ± 4.5 & - & - \\ \hline
%30 & NIH & 88.72 ± 3.23 & - & - \\ \hline
%31 & NIH & 86.9 ± 4.9 & - & - \\ \hline
%\textbf{Ours}&	AbdominalCT-1K & 88.31 ± 7.24&	5.10 ± 8.43&	1.17 ± 1.49\\ \hline
%\end{tabular}
%\label{evaluation}
%\end{table*}
%% zheyuan --> here put a table from table 1, only ct studies and put  your ct results as a comparison. It means, you create a table, and you put your results from your algorithm at the last row, and then you put other studies from table 1 where they used NIH and otherdataset. I just want you to compara your results with them. You have your ct results in table 3. you just need copy paste and organization.

\blue{We evaluated \textit{PanSegNet} against several state-of-the-art methods, including \textit{nnUNet}~(\citep{nnunet}), \textit{nnUNet-Res (size M)}, ~\textit{SSformer, SwinUNETR, MedSegDiff, SynergyNet, VNet,} (\citep{ssformer, swinunetr, medsegdiff, synergynet, vnet}) and \textit{TransBTS}, on our combined datasets.} Unlike the previous section, focusing on analyzing the domain shift between different domains, this section particularly focuses on benchmarking various models' performance. To ensure consistency, we partitioned the combined dataset into three subsets: 75\% for training, 5\% for validation, and 20\% for testing, maintaining this split across all baseline models. We repeated the experiments both for T1W and T2W scans. Quantitative results in Table~\ref{quant:bench_comp} illustrate \textit{PanSegNet's} superior performance on T1W images across all metrics. Notably, 2D CNN-based method \textit{nnUNet} generally outperformed 2D transformer-based methods such as \textit{SSFormer} and \textit{SwinUNETR}. However, \textit{MedSegDiff}, a 2D CNN-diffusion model-based method, exhibited superior performance, likely due to its utilization of diffusion models with fixed procedures and high-dimensional latent variables, enabling more comprehensive data representation. 

On T2W MRI images, similarly, \textit{PanSegNet} also demonstrated superior performance across most metrics. T2W images pose challenges for deep learning models as they are more sensitive to distinguishing the pancreas from nearby fluid-filled structures, making delineation more challenging. \blue{It is worth noting that nnUNet-Res (size M) outperformed nnUNet in 3D segmentation tasks, indicating that residual encoder available in nnUNet-Res enhances the network's ability to learn complex features and improves gradient flow during training~\citep{nnures}. nnUNet-Res comes with two more sizes (L and XL) apart from size M; however, both L and XL models require significantly more VRAM and longer training times compared to the smaller "M" configuration.} 

While \textit{SynergyNet} showed competitive performance, attributed to its discrete module capturing pancreas shape and size independently of modality. \textit{PanSegNet}'s integration of 3D convolutions with linear self-attention blocks made it more efficient and superior in accurately delineating objects and boundaries. The poor performance of VNet, a basic 3D convolution-based network, highlights the significance of \textit{PanSegNet}'s design.

\textit{PanSegNet}'s robust performance across both T1W and T2W MRI modalities collected from multiple centers underscores its generalization ability. This ability to perform well across diverse datasets from different imaging centers suggests \textit{PanSegNet}'s efficacy in real-world clinical settings, where data variability is common. Moreover, \textit{PanSegNet} achieves this generalization while remaining parameter-efficient, meaning it achieves superior results without an excessive number of model parameters. This efficiency is crucial for practical deployment, as it reduces computational resources and model complexity while maintaining high performance. Thus, \textit{PanSegNet} represents a promising solution for accurate and efficient pancreas segmentation in clinical practice.
\subsubsection{Ablation studies}
Our proposed method is built upon the nnUNet framework, utilizing nnUNet 3D as the baseline in the absence of linear attention. As shown in Table~\ref{quant:bench_comp}, our method exhibits promising performance compared to the version without linear attention. 

\subsubsection{Inference time}
To evaluate inference efficiency, we conducted 100 inference runs on standard nnUNet patch sizes. Our model, leveraging linear attention, introduced negligible additional computational overhead despite its relatively dense parameterization. All experiments were executed on an A6000 server.
\begin{table}[!ht]
    \caption{Inference time analysis for PanSegNet. We can observe that limited computational overhead is introduced by linear attention block.}
    \centering
    \resizebox{\linewidth}{!}{
    \begin{tabular}{c|c|c|c}
    \hline
        ~ & Gflops (G) & Inference Time (s) & Parameters  (M) \\ \hline
        Baseline & 549.5976 & 0.1114 & 30.97 \\ \hline
        Ours & 552.3234 & 0.1152 & 43.92 \\ \hline
    \end{tabular}}
\end{table}
\subsection{Comparison with foundational model}

Given the inherent challenges in segmenting the pancreas, the foundational model struggles in this context due to the limited availability of MRI training data. Most foundational models, such as SAM \citep{kirillov2023sam} and MedSAM \citep{ma2024medsam}, depend on input bounding boxes or points for 2D segmentation, making direct comparisons to our work difficult. \blue{Nevertheless, we have conducted the segmentation at the 2D slice level using SAM and MedSAM with generated bounding boxes from the ground truth mask as input on data from Center 3; in contrast, the bounding box information is not accessible to our model. }The results Table~\ref{quant:fundational_model}, as presented in the following table, demonstrated that even with the aid of bounding boxes, SAM and MedSAM fall short of achieving the performance levels of our model. Additionally, the data revealed that SAM and MedSAM performed better on T1 scans compared to T2 scans, likely due to the superior contrast present in T1 images and inferior to the proposed \textit{PanSegNet}.

\begin{table}[h]
\centering
\caption{Quantitative segmentation performance evaluation and Comparison with diverse state-of-art methods across multi-center T1W and T2W MRIs.}
\resizebox{\linewidth}{!}{
\begin{tabular}{c|c|c|c|c} 
\hline
\rowcolor[rgb]{0.9, 0.9, 0.9} \multicolumn{5}{c}{\textbf{Multi-center T1W MRI Pancreas Segmentation}} \\ \hline
\rowcolor[rgb]{0.9, 0.9, 0.9} \textbf{Methods} & \textbf{Modality} &\textbf{Dice (\%)}& \textbf{HD 95 (mm)}&	\textbf{ASSD (mm)} \\ \hline
\textbf{SAM}& 2D & 40.83 & 42.45 & 9.9  \\ \hline

\textbf{MedSAM}& 2D & 43.08 & 42.72 & 9.45 \\ \hline

\textbf{PanSegNet}& 3D & \textbf{81.57} & \textbf{5.64} & \textbf{1.78} \\ \hline

\rowcolor[rgb]{0.9, 0.9, 0.9} \multicolumn{5}{c}{\textbf{Multi-center T2W MRI Pancreas Segmentation}} \\ \hline
\rowcolor[rgb]{0.9, 0.9, 0.9} \textbf{Methods} & Modality &\textbf{Dice (\%)} & \textbf{HD 95 (mm)}&	\textbf{ASSD (mm)} \\ \hline

 \textbf{SAM} & 2D & 32.03 & 45.35 & 11.27 \\ \hline

\textbf{MedSAM}& 2D & 29.86 & 45.37 & 11.04 \\ \hline \hline

\textbf{PanSegNet}& 3D & \textbf{87.83} & \textbf{5.33} & \textbf{0.97}  \\ \hline
\end{tabular}}
\label{quant:fundational_model}
\end{table}

%%%%%%%%%%%%%%%%%%%
\begin{table*}[h]
\centering

\caption{ Quantitative segmentation performance evaluation and Comparison with diverse state-of-art methods across multi-center T1W and T2W MRIs.}
\begin{tabular}{|c|c|c|c|c|c|c|c|} 
\hline
\rowcolor[rgb]{0.9, 0.9, 0.9} \multicolumn{8}{|c|}{\textbf{Multi-center T1W MRI Pancreas Segmentation}} \\ \hline
\rowcolor[rgb]{0.9, 0.9, 0.9} \textbf{Methods} & Modality &\textbf{Dice (\%)}&	\textbf{Jaccard (\%)}&	\textbf{Precision (\%)}&	\textbf{Recall (\%)}&	\textbf{HD 95 (mm)}&	\textbf{ASSD (mm)} \\ \hline

 \textbf{nnUNet}& 2D & 	80.19 &	80.01  &	80.19 &	81.01 &	20.99  &	1.75 \\ \hline

\textbf{SSformer}& 2D &	78.81 &	77.01 &	76.67 &	75.89 &	23.09  & 2.54 \\ \hline

\textbf{SwinUNETR}& 2D &	76.01 &	75.21 & 73.21 &	74.11 &	27.78 &	2.98 \\ \hline

\textbf{MedSegDiff}& 2D &	83.75 &	82.11 &	81.78 &	80.99 &	18.97 & 1.56 \\ \hline

% \textbf{SAM-Base}& 2D &	\textbf{96.00} &	\textbf{92.38} &	82.55 &	83.02 &	19.01 & 1.61 \\ \hline

\textbf{SynergyNet}& 2D &	85.78 &	84.37 & 84.09 & 84.44 & 17.88  & 0.95 \\ \hline

\textbf{VNet}& 3D &	73.15 &	74.01 & 72.11 &	84.92 & 71.47  &  2.99 \\ \hline

\textbf{TransBTS}& 3D &	75.89  & 74.18 & 74.87 &	73.92 &	26.44 &	3.01 \\ \hline

\textbf{MedNext} & 3D & 80.05 & 79.99 & 83.33 & 80.02 & 17.77 & 1.67 \\ \hline

\textbf{nnFormer}& 3D & 82.11 & 83.28 & 83.23 & 81.11 & 18.45 & 1.98  \\ \hline
\textbf{nnUNet}& 3D & 80.09 & 81.29 & 83.87 & 81.98 & 18.12 & 1.79  \\ \hline
\textbf{nnUnet-Res}& 3D & 83.02 & 84.01 & 82.91 & 82.54 & 17.92 & 1.52 \\ \hline \hline

\textbf{PanSegNet}& 3D &	\textbf{86.02} &	\textbf{85.78} & \textbf{84.18} &	\textbf{84.76} &	\textbf{17.47} & \textbf{0.92} \\ \hline

\rowcolor[rgb]{0.9, 0.9, 0.9} \multicolumn{8}{|c|}{\textbf{Multi-center T2W MRI Pancreas Segmentation}} \\ \hline
\rowcolor[rgb]{0.9, 0.9, 0.9} \textbf{Methods} & Modality &\textbf{Dice (\%)}&	\textbf{Jaccard (\%)}&	\textbf{Precision (\%)}&	\textbf{Recall (\%)}&	\textbf{HD 95 (mm)}&	\textbf{ASSD (mm)} \\ \hline

 \textbf{nnUnet} & 2D & 80.96 &	81.98 &	82.01 &	83.11 &	19.88  &	1.76 \\ \hline
 
\textbf{SSformer}& 2D &	80.01 &	79.09 & 80.01 &	80.19 &	22.15 &	2.02 \\ \hline

\textbf{SwinUNETR}&	2D & 78.71 &	76.34 &	75.78 &	76.01 &	26.89  & 2.76 \\ \hline

\textbf{MedSegDiff}& 2D &	85.01 &	83.47 &	84.04 &	83.78 &	17.47 & 1.32 \\ \hline

% \textbf{SAM-Base}& 2D &	84.88 &	83.14 & 82.99 &	81.77 &	18.22  & 1.37 \\ \hline

\textbf{SynergyNet}& 2D &	\textbf{86.51} &	85.52 &	84.76 &	85.19 &	17.75  & 0.91 \\ \hline

\textbf{VNet}& 3D &	73.75 & 74.09 & 72.56	 &	73.79 &	27.19 & 3.0 \\ \hline

\textbf{TransBTS}& 3D &	76.92 &	75.18 &	75.21 &	76.88 &	27.01  & 2.79 \\ \hline

\textbf{MedNext} & 3D & 79.98 & 84.07 & 84.23 & 83.98 & 19.01 & 1.92 \\ \hline

\textbf{nnFormer}& 3D & 80.04 & 83.98 & 83.21 & 83.39 & 19.08 & 2.09  \\ \hline
\textbf{nnUNet}& 3D & 80.88 & 84.56 & 84.11 & 84.17 & 18.82 & 1.88  \\ \hline
\textbf{nnUnet-Res}& 3D &  81.11 & 85.12 & 84.91 & 84.44 & 18.47 & 1.48 \\ \hline \hline

\textbf{PanSegNet}& 3D &	{86.01} &	\textbf{86.78} &	\textbf{85.77} &	\textbf{85.88} &	\textbf{17.23}  & \textbf{0.88} \\ \hline
\end{tabular}
\label{quant:bench_comp}
\end{table*}

\subsection{Exploratory Analysis of Segmentation Quality and Volumetric Accuracy}
Current medical image segmentation relies on the region-based (Dice, F1-score) and boundary-based (Hausdorff distance, surface distance) metrics as de facto standard. While these metrics are widely used for detailed regional and boundary analysis, there is a need for clinicians to compare volumetric agreement too.  Recognizing the clinical relevance of volumetry, we utilize relative volume prediction error (vpe) to directly assess the accuracy of volume predictions derived from segmentation tasks. In~\citep{hussain2021vest}, the authors estimated vpe from the Dice Coefficient based on the assumption that the Dice coefficient is close to 1 and segmentation performs well. In this exploratory analysis, we integrate a theoretical analysis and empirical validation across diverse datasets under general cases by following~\citep{hussain2021vest} analysis for the first time a challenging organ like pancreas where dice is far from 100\% due to the complexity of the organ's segmentation procedure. Our findings below highlight the critical role of incorporating volumetric prediction accuracy into segmentation evaluation. This approach empowers clinicians with a more nuanced understanding of segmentation performance, ultimately improving the interpretation and utility of these metrics in real-world healthcare settings.

\subsubsection{Theory proof}
For any given segmentation task~\citep{hussain2021vest}, one can calculate both the Dice coefficient ($dice$, between 0 and 1) and the relative volume prediction error (vpe), which is a direct evaluation metric for volume statistics and can be represented as:
\begin{equation}
    vpe = \frac{volume_{predict} - volume_{real}}{ volume_{real}}.
\end{equation}

Given the $p_i, q_i$ represent ground truth and prediction masks (in binary format-0 for background, 1 for foreground), dice coefficient ($dice$ from now on for simplicity) between these masks can be calculated as:
\begin{equation}
    dice = \frac{2\sum_i p_i q_i}{ \sum p_i + \sum q_i}.
\end{equation}
We observe that, 
\begin{equation}
    \sum_i (p_i q_i) \leq min(\sum p_i, \sum q_i),
\end{equation}
as $\sum_i (p_i q_i)$ is an overlap region between masks and is measured by the number of pixels (voxels). Thus, we can replace $dice$ equation with this new bound as:
\begin{equation}
    dice \leq min(2\frac{\frac{\sum p_i}{\sum q_i}}{1 + \frac{\sum p_i}{\sum q_i}}, 2\frac{1}{\frac{\sum p_i}{\sum q_i} + 1}).
\label{eq:initial_dice}
\end{equation}
Recalling that $volume_{predict}=\sum p_i, volume_{real}=\sum q_i$, the relative volume prediction error can be obtained as: 
\begin{equation}
    vpe = \frac{\sum p_i}{\sum q_i} - 1.
\label{eq:vpe_define}
\end{equation}
In the original $dice$ formulation (Eq.~\ref{eq:initial_dice}), now we replace the variable $\frac{\sum p_i}{\sum q_i}$ with $vpe+1$ and obtain
\begin{equation}
    dice \leq min(\frac{2}{2+vpe}, \frac{2+2vpe}{2+vpe}).
\end{equation}

Note that $vpe \geq -1$ from Eq.\ref{eq:vpe_define}, thus, by adding constant 2 into both sides of the equations, we can conclude that $2+vpe\geq 1$. Now, we reconsider denominators of the $dice$ bound equation above (Equation 6), and note that the minimum of the two is the $\frac{2}{2+vpe}$; hence,
\begin{gather}
\label{eq:full_compare}
    dice \leq \frac{2}{2+vpe} \implies vpe \leq \frac{2}{dice}-2 (\geq 0), \\
    dice \leq \frac{2+2vpe}{2+vpe} \implies vpe \geq \frac{2}{2-dice} -2 (\leq 0).
\end{gather}
This equation defines the \textbf{upper} and \textbf{lower} boundary for volume prediction error (vpe) associated with the basic dice coefficient. It is worth noticing that the equation derived in ~\citep{hussain2021vest} is actually \textbf{upper} boundary when the Dice coefficient is close to 1 and not suitable for the general cases. In most cases, we care about the mean absolute vpe as $avpe_{m} = \frac{1}{N} \sum |vpe|$. From the HM-GM inequality~\citep{sedrakyan2018algebraic}, one can observe that
\begin{gather}
    \frac{1}{2}(\frac{1}{dice} + \frac{1}{2-dice}) \geq \frac{2}{dice + 2 - dice} = 1, \\
    \implies 2-\frac{2}{2-dice} \leq \frac{2}{dice}-2.
\end{gather}
This leads to the fact that $|avpe|$ is bounded as:
\begin{gather}
  %  |vpe| \leq \frac{2}{dice} - 2 \\
    avpe_{m} = \frac{1}{N} \sum |vpe| \leq \frac{1}{N} \sum (\frac{2}{dice} - 2 ) \leq \frac{2}{\frac{1}{N}\sum dice} - 2, \\
    \implies avpe_{m} \leq \frac{2}{dice_m} - 2.
    \label{eq:dice_volume_corr}
\end{gather}

\begin{figure*}
\centering
\includegraphics[width=1\textwidth]{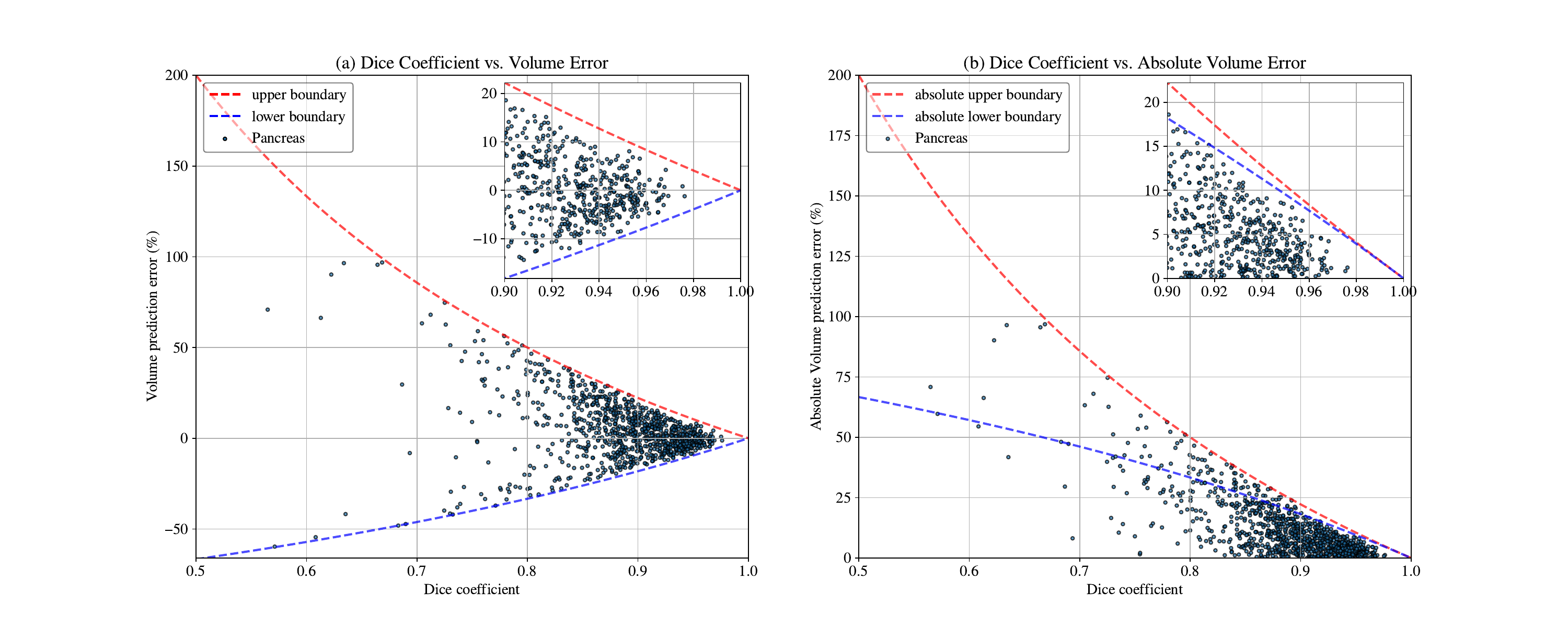}
\caption{Upper and lower bound information for dice vs volume error (left) and dice vs absolute volume error (right) are given, respectively. For example, for a 94\% dice coefficient, 10\% volume error is quite plausible while this is not the case for other organs in general,  but pancreas.}
\label{volumebounded}
\end{figure*}

Figure~\ref{volumebounded} illustrates the relation between volume error and dice coefficient for pancreas segmentation in particular. For any given dice score, an estimated (or plausible) pancreas volume error is shown with red and blue markers (upper and lower bounds). These upper and lower bounds in volume errors highlight the clinically useful dice scores by looking at the pancreas volume prediction error. For instance, for population and diagnostic studies 10-20\% volume error is often found plausible, indicating a dice score close to 90\% is acceptable at the clinics. On the other hand, a higher level of precision may be required for surgical planning in a different clinical context, such as in pancreas transplantation or pancreatic resection. In these scenarios, an acceptable error percentage of around 3\% to 5\% might be necessary to ensure accurate preoperative assessment and intra-operative guidance. In this scenario, approximately 96\% dice will be required, which is the limitation of all available algorithms currently, while ours is the closest one to the desired level, perhaps the only one in the literature so far.

\section{Discussion and Concluding Remarks}
This study aimed to develop and evaluate an accurate and generalizable deep learning method for pancreatic segmentation, \textit{PanSegNet}.  It transcends the limitations of existing methods and offers \textit{multi-modality compatibility} and enhanced \textit{generalizability}. 

\textit{PanSegNet} was tested on the first multi-center, multi-modal, large-scale MRI pancreas dataset for pancreas segmentation. It is equally effective on CT, T1W, and T2W MRI scans and achieves consistently high Dice coefficients and low Hausdorff distances (HD95), serving as the state-of-the-art results in the literature. \textit{PanSegNet} demonstrated a robust cross-platform performance across diverse datasets from multiple healthcare centers. \textit{PanSegNet} and its models will be publicly available to foster collaboration and accelerate research progress. 

The limited availability of publicly available pancreatic MRI datasets has hampered the development of robust automated segmentation methods compared to the abundance of CT data. The only publicly available pancreas dataset in MRI is AMOS~\citep{ji2022amos}, providing 40 T1W MRI scans (no T2W) with ground truths available~\citep{ji2022amos,yao2020advances}. This scarcity is understandable due to the challenges of collecting and annotating MRI data~\citep{scialpi2016pancreatic}. This study addresses the critical scarcity of pancreatic MRI data by introducing the first-ever multi-center, multi-modality dataset encompassing 385 T1W and 382 T2W scans with high-quality annotations.

\textit{PanSegNet} provided excellent Dice coefficients for T1W and T2W MRI segmentation (Figures 3 and 4). We also observed that our external validations showed domain shifts (Centers \#3, \#4, \#5). Similar shifts happened when CT data was chosen from external data sets (e.g., AMOS, WORD, and BTCV) (Figure~\ref{domainshift}). Collecting larger and more diverse imaging data can be a potential solution for further improving the segmentation model. However, large-scale data gathering is not only costly but there are also ethical and regulatory considerations to address.  Without trying to acquire million-scale MRI and CT scans, one alternative solution towards more generalizable segmentation might be domain generalization approaches~\citep{dg} or test-time domain adaptations~\citep{ttda}. Current domain generalization studies explore techniques to improve a model's ability to generalize knowledge learned from a specific domain to unseen domains. These methods include data augmentation, adversarial domain training, and domain-invariant feature learning~\citep{dg,dgzheyuan}. By incorporating these techniques into our segmentation model, we may mitigate the performance drop experienced during domain shifts. Studies in this field are limited.

The highly accurate segmentation and volumetry achieved by the \textit{PanSegNet} could benefit the clinical evaluation of pancreatic diseases (e.g., pancreatic cyst follow-up, chronic pancreatitis or diabetes mellitus). This automated volumetry would allow quantitative analysis in diagnosis and follow-up in response to drug therapies. A recent multi-institutional, multi-vendor study analyzed and reported quantitative and semi-quantitative parameters of the pancreatic parenchyma, including T1W signal (“T1 Score”), arteriovenous enhancement ratio (AVR), pancreas volume and pancreas diameter~\citep{tirkes2019magnetic}. The authors also proposed two multi-parametric composite scores to obtain higher diagnostic performance for CP~\citep{tirkes2023diagnosis}. These multi-parametric scores were SQ- score, AVR venous, and pancreas volume). AUCs for Models A and B were higher than using individual parameters (0.92 and 0.93, respectively).   

Accurately measured pancreas volume can offer valuable indices for risk stratification of diabetes mellitus (DM). While the precise dynamics of pancreas volume decline in DM remain unclear, studies have consistently shown a reduced pancreas size in patients with newly diagnosed type 1 diabetes (T1D)~\citep{sasamori2018analysis, campbell2016influence, williams2012pancreatic}. A recent study investigated the temporal dynamics of pancreas volume in children with recent onset T1D and individuals without diabetes utilizing quantitative MR techniques~\citep{virostko2019pancreas}. At enrollment, the pancreas volume index was lower in patients with recent onset T1D than in controls (median 0.600 mL/kg in T1D vs 0.929 mL/kg in controls; $p < 0.001$). MRI measurements of the pancreas at 6 and 12 months after diagnosis of T1D revealed a continuing decline in pancreas volume index (0.6\% per month in T1D patients compared to the control cohort, p=0.001).

Our study has a few limitations. First, the \textit{PanSegNet}'s adaptability and performance in broader clinical applications could be a potential limitation as only pancreas is considered as the main organ herein. Second, limitations may arise from the quality of MRI scans or the presence of artifacts. While we already pre-process MRI scans for segmentation, more sophisticated image harmonization methods can be adapted for improved segmentation results from varying sequence differences. Furthermore, we did not have the opportunity to validate our reported results in a prospective clinical setting; however, our model is made publicly available for use in such contexts.

\section*{Acknowledgments}
This work is supported by NIH funding: R01-CA246704, R01-CA240639, U01-DK127384-02S1, and U01-CA268808.

%%Harvard
\bibliographystyle{model2-names.bst}\biboptions{authoryear}
\bibliography{refs}

\end{document}